\documentclass[twoside,11pt,letter,onecolumn]{article}

\newcommand{\Milexiste}{\ensuremath{\exists}}

\newcommand{\Mnon}{\ensuremath{\lnot}}
\newcommand{\Munion}{\ensuremath{\cup}}
\newcommand{\Minter}{\ensuremath{\cap}}

\newcommand{\round}[1]{\ensuremath{\textbf{[}#1\textbf{]}}}
\newcommand{\cround}[1]{\ensuremath{\textbf{\{}#1\textbf{\}}}}
\def\sqw{\hbox{\rlap{\leavevmode\raise.3ex\hbox{$\sqcap$}}$%
\sqcup$}}
\def\sqb{\hbox{\hskip5pt\vrule width4pt height6pt depth1.5pt%
\hskip1pt}}
\def\qed{\ifmmode\hbox{\hfill\sqb}\else{\ifhmode\unskip\fi%
\nobreak\hfil
\penalty50\hskip1em\null\nobreak\hfil\sqb
\parfillskip=0pt\finalhyphendemerits=0\endgraf}\fi}
\def\cqfd{\ifmmode\sqw\else{\ifhmode\unskip\fi\nobreak\hfil
\penalty50\hskip1em\null\nobreak\hfil\sqw
\parfillskip=0pt\finalhyphendemerits=0\endgraf}\fi}
\newcommand{\kw}[1]{\emph{#1}}

\def\A{\mathcal A}

\def\H{\mathcal H}

\def\O{\mathcal O}

\def\S{\mathcal S}

\def\CC{\mathbb C}

\def\NN{\mathbb N}

\def\QQ{\mathbb Q}
\def\RR{\mathbb R}

\def\ZZ{\mathbb Z}

\def\pp{\vect{p}}

\newenvironment{Tproof}{{\bf Proof:} }{ $\sqw$
\vspace{0.33cm}

}
\newenvironment{Tproofsketch}{{\bf Proof(sketch):} }{ $\sqw$
\vspace{0.33cm}

}

\usepackage{fancyvrb}
\author{Bertrand Nouvel \footnote{(\texttt{<bertrand.nouvel@ens-lyon.fr>} -- This author is supported by the french television channel TF1)}, \'Eric R\'emila \footnote{(\texttt{<eric.remila@ens-lyon.fr>})}}
\title{Incremental and Transitive Discrete Rotations}
\date{UMR CNRS - ENS Lyon - UCB Lyon - INRIA 5668\\Laboratoire de l'Informatique du Parall\'elisme\\ \'Ecole Normale Sup\'erieure de Lyon\\46, All\'ee d'Italie 69364 LYON CEDEX 07 - FRANCE}

\newtheorem{definition}{Definition}
\newtheorem{lemma}{Lemma}
\newtheorem{proposition}{Proposition}

\usepackage[dvips]{graphics}

\usepackage{multicol}

\usepackage{amssymb}

\usepackage{stmaryrd}

\usepackage{algorithmic}
\usepackage[dvips]{graphics}

\usepackage{url}
\usepackage{psboxit}
\usepackage{enumerate}

\usepackage[left=1in,top=1in,right=1in,bottom=1in,nohead,nofoot]{geometry}

\begin{document}

\maketitle

\begin{abstract}
A discrete rotation algorithm can be apprehended as a parametric map $f_\alpha$ from $\ZZ[i]$ to\ $\ZZ[i]$, whose resulting permutation ``looks like'' the map induced by an Euclidean rotation. For this kind of algorithm, to be incremental means to compute successively all the intermediate rotated copies of an image for angles in-between 0 and a destination angle. The discretized rotation consists in the composition of an Euclidean rotation with a discretization; the aim of this article  is to describe an algorithm which computes incrementally a discretized rotation. The suggested method uses only integer arithmetic and does not compute any sine nor any cosine. More precisely, its design relies on the analysis of the discretized rotation as a step function: the precise description of the discontinuities turns to be the key ingredient that will make the resulting procedure optimally fast and exact.
A complete description of the incremental rotation process is provided, also this result may be useful in the specification of a consistent set of definitions for discrete geometry.
\end{abstract}

{\bf keywords : } \textit{Discrete Rotations}, \textit{Discrete Geometry}, \textit{Computational Geometry}.
\section{Introduction}

The translation of the fundamental concepts of the Euclidean geometry into $\ZZ^n$ comprises the field of discrete geometry. As this theory of geometry is particularly suitable for combinatorial images and other data manipulated by computers \cite{KR2004}, it would be interesting to provide a set of efficient algorithm 
for this theory that uses only integer-arithmetic; as this was suggested in \cite{reveilphd}.

Several attempts have been realized by various authors that wished to 
deliver back some properties of the Euclidean rotation to the 
discretized rotations widely used in computer graphics. A review of various attempts 
may be found in \cite{andresphd}. 






In this paper, we present an algorithm which is \emph{incremental}: it successively computes all the rotated images according to the an increasing sequence of angles (starting from 0 to $2\pi$). Notice that the set of rotated images is finite on a finite picture, this allows practically to compute all the intermediate rotated images.
Moreover, the suggested procedure is \emph{sound} and \emph{accurate}: it returns exactly the same results as the result provided by the discretized for the same angle. The procedure does not use any sine nor any cosine, thus there is no precision problem due to the floating point arithmetic.
Also, the algorithm is \emph{fast}: to compute incremental rotations the algorithm computes only $\O(m^3-log(m))$ operations, instead of $\O(m^5)$ as need the naive algorithm. 
For incremental rotations, the complexity of this algorithm if it uses pre-calculated tables, $\O(m^3)$ is optimal: The algorithm updates only the necessary pixels and only consider the necessary angles.
Finally, due to the fact the algorithm uses 
configurations that can be stored with very few states on the plane, we believe it is a good candidate for parallelization.


After a brief review of the motivations, and after the essential preliminary definitions, we proceed to a characterization of the discontinuities of the rotation process. Indeed, we will explain how to code the angles where the discontinuities happen. Also, with integer arithmetic only, we will learn to perform essential operations on the encoded angles. Naturally, a few technical lemmas are required to set up all this framework properly. Once this has been set, we will analyze the alterations that occur in the configuration at the discontinuities. Strengthened by previous results, we will then be ready to build the incremental
 discretized rotation procedure. The last section will be dedicated to various extensions and miscellaneous details related 
to the theory that may lead to a better understanding of the 
discretized rotation process.

\section{Groundwork}

The first sections present the fundamental ideas, definitions and lemmas that matter to fully understand the algorithm. In this section, we review the motivations and precise some vocabulary.

\subsection{Motivations}

The history of discrete geometry begins with the common will 
to give birth to an algorithmic theory of the geometry in the discrete spaces
that would be consistent with the Euclidean geometry.
We believe that a unified theory would provide a better understanding of both universe. Discrete rotations comprise the famous examples that have strengthened the idea that discrete and continuous spaces may be radically different. A review of the differences can be found in the prologue of \cite{medam}.

More recently, discretized rotations have been an important issue in water-marking community. Water-marking algorithms that were robust under rotation were sought for by  various teams the community; and it supports many discussions. More generally, the problem of finding algorithms for classification and recognition of patterns that are robust under discrete rotations is still a not-trivial issue in the conception of pattern related algorithms. 
The algorithm presented use similar principle has the one suggested by \cite{rotmatch04}; the basic idea is to compute a table of the discontinuities of the discrete rotation process, and to use this table for the rotation. Our paper provides a more accurate description of the discontinuities, as-well as a way to compute the rotation  using integer-arithmetics only.

The algorithm was designed to comprise the field of discrete geometry: our main problem is to provide a comprehensive and simple definition  of algorithm for the discrete rotation, independantly of any specific usage, and that can be computed efficiently.

The usual discretized rotation algorithm is a bit chaotic when iterated:
This is well illustrated by the following story: It has been implemented a physical simulation of the solar system which for specific reasons was coded with fixed precision. When the system has been tried, its inventors were surprised that it first conclusion was to predict a fatal collision in-between the earth and the sun in less than ten years; naturally there was a ``bug'', and it was imputable to composition of rounding errors during the rotation process. More generally, decades of computer programmers have known by experience that they should avoid to compose rotations: The accumulation of the resulting errors may produce an unwelcome result. Some aspects of the dynamical system that is formed by the iterated action of a discretized rotation have already been studied; see \cite{poggiaspalla} for iteration of $\pi/4$-rotations.

\subsection{Conventions}
We work in the {\em complex plane} $\CC$, where $\ZZ[i]$ denotes the set of Gaussian integers.
Let $m$ be a positive integer. We denote by $\ZZ[i] \vert_m$ the set of 
Gaussian integers whose module is at most $m$; $\ZZ[i] \vert_m = \{ z 
\in \ZZ[i] , \vert z \vert \leq m \}$.
Real and imaginary parts are denoted $\Re(z)$ and $\Im(z)$.
Let $x$ be a real number.
We recall that the floor function $ x \mapsto \lfloor x \rfloor$ is 
defined
as the greatest integer less or equal to\ $x$.
The \kw{rounding function} is defined as : $\round{x}=\lfloor x + 0. 5 
\rfloor$; we also define the map $x \mapsto \cround{x}$ by 
$\cround{x}=x-\round{x}$.
These maps are extended to complex numbers, by applying them 
independently on the real part and on the imaginary part.
Let $\H$ be set of complex number that have a semi-integer coordinate, in the general case $\H$ denotes the set of discontinuity points of the operator $x\mapsto \round{x}$ ( in other terms $\H=(\RR\times (\ZZ+\{\frac12\}))\cup ((\ZZ+\{\frac12\})\times \RR)$ ).


Let $\alpha$ denote a angle, i.e. an element of $\A=\RR/(2\pi\ZZ)$.
The {\em Euclidean rotation} $r_\alpha$ is the bijective isometry of $\CC$: $z 
\mapsto ze^{i\alpha}$.
\emph{The discretized rotation} $\round{r_\alpha}$ is precisely defined 
as the successive computation of the Euclidean rotation of angle $\alpha$ and of the discretization operator $z \mapsto \round{z}$.
 Thus, for each $z$ of $\CC$, $\round{r_\alpha(z)} = 
\round{z e^{i \alpha}}$. Remark that
 $\vert \round{r_\alpha(z)} \vert \leq \vert z \vert +\frac{\sqrt{2}}2$, and that
 $\vert \round{r_\alpha(z)} - r_\alpha(z) \vert \leq \frac{\sqrt{2}}2$.


A \emph{configuration} is a mapping from $\ZZ[i]$ to $\ZZ[i]$. Let $m$ 
be a positive integer. A configuration  $C$ such that for all $z \in \CC$, $ \lim_{n\mapsto \infty}   \frac{C(\round{2^n\,z})}{2^n}-(e^{i\alpha}z)  \to 0$ is called an $\alpha$-\emph{rotation map}.  
Thus, given a real $\alpha$, the discrete 
rotation $\round{r_\alpha}$ is a configuration.
A partial configuration of radius $m$ is a mapping from $\ZZ[i] 
\vert_m$ to $\ZZ[i]$. Each configuration induces a partial 
configuration. In this paper, we work on on partial configurations 
(denoted by $\round{r_\alpha} \vert_m$) induced by discretized 
rotations. Precisely, we study the mapping ${\rho}_m$ : $\alpha \mapsto 
\round{r_\alpha} \vert_m$.
Since $\ZZ[i] \vert_m$ is finite and $\vert \round{r_\alpha(z)} \vert 
\leq \vert z \vert +1$, the set
$ \{ \rho_m (\alpha) , \alpha \in \RR \}$ is finite.

In this paper, we produce an \emph{exact incremental rotation 
algorithm} which, given an integer $m$ successively produces all the 
values $\rho_m (\alpha)$ for $\alpha \in [0, 2 \pi ]$, in the order 
where they are reached (moving from 0 to $2\pi$). Informally, it is a 
"video" algorithm which exhibits the successive configurations 
obtained along the rotation.

\section{Hinge Angles}

\subsection{Definitions}

\begin{figure}[tbp]
 \centering
$ \begin{array}{c}
 \resizebox{6.66cm}{6.66cm}{\includegraphics{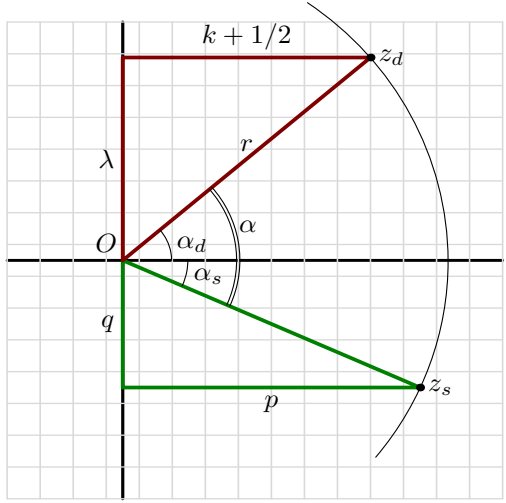}}\\
 \end{array}$
 \caption{An hinge angle and its generating pair for $\alpha=\alpha(9,-4,7)$.}
 \label{fig:spctriangle}
\end{figure}

The principal element that have influenced the design of the algorithm 
 is a precise study the function $\rho_m$. 
We shall prove that $\rho_m$ is a stair function(piecewise constant function). Thus, a precise study of its discontinuity steps gives the ability recover the whole function. The discontinuities also correspond to the only updates of the configuration stored in the algorithm we will describe. 

\begin{definition}
An angle $\alpha$ is a {\em hinge angle} if there exists a {\em source 
point} $z_s$ in $\ZZ[i]$ such that
the {\em destination point} $z_d=z_se^{i\alpha}$ has a (proper) semi-integer 
component (i.e. $z_d\in \H$) . For each 
hinge angle, the source point and the destination point form a {\em generating pair}.
\end{definition}

On a topological point of view, the hinge angles are  the discontinuity points 
of the map $\alpha \mapsto \round{r_\alpha}$.

It is immediate from the definition that an angle is a hinge angle if and only if there exists integers $p,q,k$ such that $2\,q \cos(\alpha) + 2\,p \sin(\alpha) = 2\,k+1$. 

The value $\vert z_s(\alpha) \vert^2$ is called the \emph{order} of  $\alpha$. We can note that if $(z_s,z_d)$ generates $\alpha$ then $(\bar{z}_s,\bar{z}_d)$  generates $-\alpha$.Note, also that if $(z_s,z_d)$ generates $\alpha$ then $(i^Qz_s,i^Qz_d)$ generates the same angle, for any $Q\in \{0,1,2,3\}$. 
We also define $S_\alpha$  the set of source points of $\alpha$: 
$S_\alpha=\{z \in \ZZ[i]| e^{i\alpha}z \in \H\}$ 

\def\mod{\ensuremath{\mathrm{\ mod \ }}}

\subsection{Fundamental Lemmas}

The \kw{Pythagorean angles}, as seen in \cite{merrajlater} or \cite{VOSS93}, are such that $\alpha=\arctan(a/b)$ where $a$ and $b$ are issued from a Pythagorean triple $(a,b,c) \in \NN^3$ (such that $a^2+b^2=c^2$). $c$ will be called the \emph{radius} of the Pythagorean angle.  An angle is Pythagorean if and only if its cosine and sine are rational. 

\begin{lemma}
\label{thm:spcpythdisj}
Hinge angles and Pythagorean angles form disjoint sets. 
\end{lemma}

\begin{Tproof} %
Assume that there exists an angle which is both Pythagorean and hinge. By definition, as it is a hinge angle, there exists a Gaussian integer $z_s$ of $\ZZ[i]$ which is transformed by rotation in a point of $\H$. %
 %
We can easily check that: $\cround{\Re(e^{i\alpha}z_s)}=\cround{\Re(z_d)}=\frac{1}{2}$. More exactly,
$\Re(z_d)=\Re(z_s) \,\cos(\alpha) - \Im(z_s)\,\sin(\alpha)$. If $\alpha$ is Pythagorean, $cos(\alpha)=\frac{a}{c}$ and $sin(\alpha)=\frac{b}{c}$, where $(a,b,c)$ is a primary Pythagorean triple: i.e. belongs to $\ZZ$ and $a^2+b^2=c^2$, $gcd(a,b,c)=1$.
$\cround{\Re(z_d)}$ can be written as $\frac{n}{c}$ where $n$ as an integer, while $c$ is  odd. But this is contradicting: $\cround{\Re(z_d)}=\frac12$
\end{Tproof} 

A corollary of this lemma is that for a hinge angle, at least one of $\cos(\alpha)$ and $\sin(\alpha)$ is an irrational quadratic number.

Let $p,q,k$ be a triple of integers, such that $p^2+q^2>|k+\frac12|$, we state $r^2=p^2+q^2$, and $\lambda=\sqrt{r^2-(k+\frac12)^2}$ we define the angle  $\alpha(p,q,k)$ by the equality $e^{i\alpha(p,q,k)}(p+qi)=k+\frac12+\lambda\,i$. Thus for this angle $((p+qi), (k+\frac12+\lambda\,i))$ forms a generating pair.


\begin{lemma}[Coding of Hinge Angles]
\label{thm:onedir}
Let  $(p,q,k)$ and $(p',q',k')$  be  two distinct generating triples 
such that $\alpha(p,q,k)=\alpha(p',q',k')$.
Then $\det\left[ \begin{array}{cc} p & q \\ p' & q' \end{array}\right]=0$.

\end{lemma}

\begin{Tproof} %
The proof is straightforward: 
Let $\alpha=\alpha(p,q,k)$, thus
$2p \sin(\alpha) + 2q \cos(\alpha)=2k+1$, and similarly;
$2p' \sin(\alpha) + 2q' \cos(\alpha)=2k'+1$.
Consider these two equations as a linear system of $\cos(\alpha)$ and $\sin(\alpha)$.
If $\det(\left[ \begin{array}{cc} p & q \\ p' & q' \end{array}\right])\neq 0$ then $\cos(\alpha)$ and $\sin(\alpha)$ 
have to be rational, and the angle have to be Pythagorean.
But this would contradict lemma \ref{thm:spcpythdisj}.
Therefore $(p,q)$ and $(p',q')$ are colinear.
\end{Tproof} %

The lemma implies that $arg(z_s(\alpha))\mathrm{\ mod\ } (\pi/4)$ and $arg(z_d(\alpha)) \mathrm{\ mod\ } (\pi/4)$ are uniquely defined for an angle $\alpha$. Thus, they are important characteristics of the hinge angle. 
An immediate corollary of Lemma $\ref{thm:onedir}$ is that all
generating pairs of $\alpha(z_s,z_d)$ are necessarily of the form $(k\,z_s,k\,z_d)$ or of the form $(i\,k\,z_s,i\,k\,z_d)$ with $k\in\RR$.


For a triple $(p,q,k)$ that generates an angle $\alpha$,$(p,q,k)$ is called \emph{primary} if 
it is minimal  $p^2+q^2$  among the generating triples of $\alpha$. Obviously , there exists for each hinge angle a unique primary generating triple.

\begin{lemma}[Primary Generating Triple]
\label{thm:primtriple}
Let  $(p, q, k)$ be the primary generating triple of an angle $\alpha$;
the set of generating triples of $\alpha$ is $ \{ ((2n+1)p, 
(2n+1)q, (2n+1)k + n  ) \mathrm{,\ for\ } n \in \ZZ\}$.
\end{lemma}

\begin{Tproof} 
It is evident that if $z_s=(p+qi) \in \S_\alpha$ is the source point issued from the primary generating triple $(p,q,k)$
then $(2n+1)z_s \in \S_\alpha$, since $\cround{\Re{(2n+1)e^{i\alpha}z_s}}=\frac12$
for any $n$ in $\ZZ$.
Also we can notice that for any $n\in \ZZ$, 
this is another generating triple $((2n+1)\,p,(2n+1)\,q,(2n+1)\,k+n)$
of the same angle. It is also evident that for any $n',n'' \in \ZZ$, $(2pn',2qn',n'')$ 
cannot be a generating triple of $\alpha$ since $e^{i\alpha}(p+qi)\in \H$ implies that $e^{i\alpha}(2\,pn'+2\,qn'i)  \notin \H$. 
Now, we can notice $\beta p + \beta q i$ with $\beta \in \RR \setminus \ZZ$   cannot be a source  for arithmetical reasons:  $\beta(p+qi) \notin \ZZ[i]$.
Now, assume there exists $\frac{r}s \in \QQ \setminus \ZZ$ and $k'\in \ZZ$ such that 
$(\frac{r}s p,\frac{r}s qi, k')$ is a generating pair of $\alpha$,
this would implies that $\frac{r}s(p+qi)$  is a Gaussian integer and thus $gcd(p,q)$  should be divisible by $s$ (as $r$ is not). Moreover $\Re(e^{i\alpha}(\frac{r}sp+\frac{r}sqi))=\frac{2k'+1}2$. This can only happen if $2k'+1$ is also divisible by $s$.
Now $p,q,2k+1$ are all divisible by $n$, thus the generating pair is not primary, and this contradicts our hypothesis. Thus, $\beta(p+qi) \notin S_\alpha$  for   $\beta \in \RR \setminus \ZZ$.
Thus, consequently to previous lemma and to these points, there is no other generating triples for the angle $\alpha$ than the listed ones.
\end{Tproof} 
\subsection{Main Properties}

From the previous lemmas, we can state the following properties that are useful for the rotation algorithm.
\label{sect:hingecount}

\begin{proposition}
Any hinge angle can be uniquely described by its primary generating triple.
\end{proposition}
This proposition is actually a corollary of Lemma \ref{thm:onedir} and \ref{thm:primtriple}.

\begin{proposition}
The number of hinge angles of order at most $m$ is lower than $8m^3$.
\end{proposition}
This is easily proven by the fact that $|p|<m,|q|<m,|k+\frac12|<m$

Upper bounds on the number of possible hinge angles can be found in \cite{rotmatch04}. This formula can be slightly refined by using $r_2(k)$ that  represents the number of decomposition of an integer as the sum of two squares. The upper bound on the number of possible hinge angles is then:
$\#(\A_\H|_m)\leq \sum_{i=1}^{m} r_2(i) \lfloor \sqrt{i}-\frac12 \rfloor \leq \left(\sum_{i=1}^{m} r_2(i) \right) \lfloor \sqrt{m}-\frac12 \rfloor \leq (2m)^3 $.
Finally note that there is twice more rotation maps in $\rho_m(\alpha)$ than the number of hinge angles in $\A_\H|_m$.


\begin{proposition}
The elements of $S_\alpha$ forms  4 rays: 
$\displaystyle S_\alpha = \Munion_{Q \in \{0,1,2,3\}} \{ (2l+1)i^Q (p+qi), l \in \NN \}$
\end{proposition}

This proposition is another consequence of  \ref{thm:primtriple}.


\section{The Algorithm}

Schematically, the algorithm is structured in two main parts: 
During this first period of the algorithm, it starts with the enumeration of all the hinge angles whose order is smaller than the maximum order of the points
of the picture. The angles are not represented by floating point values but 
 only via their associated triples (three small integer numbers). 
This encoding of the angles provides a way to recover all the necessary information required to transform one image into the next image during the incremental rotation process.
 This first part of the algorithm, can be seen as a process similar to the generation of a  sine table and, ideally, it could be done once for all.

The second part of the process consists in a loop through the cycle of hinge angles. This loop applies successively the small transformations that are required to pass from $\rho_m(\alpha)$ to $\rho_m(\alpha')$ (\ where $\alpha'$ is the successive representative angle\ ). Practically, the algorithm stores the map $\rho_m(\alpha)$, this allows to know directly the position of the image through the transformation $\round{r_\alpha}|_m$. However, for various usages, such rotations in cellular automata, the user should may prefer variant encoding of the transformation such as rotation configurations (maps from $\ZZ[i]$ to a finite set) \cite{medam}. The principle of this algorithm is translatable on these kind of configurations.

\subsection{Enumerate and Sort Hinge Angles}

To enumerate all hinge angles in $\A_\H|_m$, it sufficient to enumerate all the Gaussian integers whose module smaller than $m$ (as source point), and to consider all possible semi-integers whose absolute value is smaller than $m$. This requires a time of $\O(m^3)$. 
Some angles will be enumerated twice, but  the duplicates can be identified and removed 
during the sorting process.

The usual sorting algorithms handle $n$ elements in $\O(n\log(n))$ operations, if we
 provide them a constant time comparison operator. The goal of this section is to 
explain how to compare two hinge angle via their generating triple coding. This comparison needs constant time.
Using  a quicksort on generated triples, the enumeration and the sorting of the hinge angles can therefore be computed in $\O(m^3log(m))$.





\begin{lemma}[Integer-Based Comparison]
\label{thm:trxthm0}
Let $\alpha \in \A_\H$ which is associated to the triple $(p,q,k)$ and
$\alpha' \in \A_H$ which is associated to the triple $(p',q',k')$,
it is possible to decide, in constant time, whether $\alpha< \alpha'$ using by knowing
$p,q,k,p',q',k'$.
\end{lemma}

\begin{Tproofsketch}
With the notations previously introduced we have:
$\cos(\alpha)=(p(k+\frac12)+q\lambda)/(p^2+q^2)$ and $sin(\alpha)=(p\lambda-q(k+\frac12))/(p^2+q^2)$. 

The key-argument is that the numbers used in the expression of $\cos(\alpha)$ and $\sin(\alpha)$ are all integer, except $\lambda$, but $4 \lambda$ is a square root of an integer, thus the other equations will simplify.

Thus the signs of $\cos(\alpha)$ and $\sin(\alpha)$ can be easily computed, reducing to the problem to the sign of an integer inequality that involves only usual operations. Hence, the ``quadrant'' of $\alpha$ can be computed with integer arithmetic-only.

If $\alpha$ and $\alpha'$ belongs to the same quadrant then we have to compare
$\cos(\alpha)$ and $\cos(\alpha')$. This can also be reduced to determining the sign of an integer expression, that can be computed  using only 
integer arithmetic.
\end{Tproofsketch}

\subsection{Moving from $\round{r_{\alpha^-}}$ to $\round{r_{\alpha^+}}$}

Let $\alpha \in \A_\H$, we now know that exists  $z_s(\alpha) \in \ZZ[i]$ and $\Re(z_d(\alpha))+\frac12 \in \ZZ$ and also $\Im(z_d(\alpha)) > 0 $. 
Due to the trigonometric orientation, and to the chosen discretization operator,  $\displaystyle \lim_{{\alpha'} \mapsto \alpha, {\alpha'}< \alpha} 
\round{r_{\alpha'}}(z_s) = \round{r_{\alpha}}(z_s)$ while $\displaystyle \lim_{{\alpha'} \mapsto \alpha, {\alpha'}> \alpha} \round{r_{\alpha}}(z_s) = (\round{r_{\alpha}}(z_s)-1)$. However, we can also notice that $\displaystyle \lim_{{\alpha'} \mapsto \alpha, {\alpha'}< \alpha} 
\round{r_{\alpha'}}(-z_s) = (\round{r_{{\alpha}}}(-z_s)+1)$ while $\displaystyle \lim_{{\alpha'} \mapsto \alpha, {\alpha'}> \alpha} \round{r_{\alpha'}}(-z_s) = \round{r_{\alpha}}(-z_s)$. These results can be summarized in the following theorem:

\begin{proposition}
\label{thm:movea}
$\forall Q \in \{0,1,2,3\}, \displaystyle \lim_{ \epsilon\mapsto 0, \epsilon>0} \round{r_{\alpha+\epsilon}}\left((z_si^Q) - \round{r_{\alpha-\epsilon}}(z_si^Q)\right)=i^{Q+2}$
\end{proposition}

Let $\psi_{z_0}$ denote the map  
such that if $z$ is in $ (z_0 i^Q (2n+1))$, with $n,Q \in \NN$, then 
$\psi_{z_0}(z,p)=p+i^{Q+1}$ else $\psi_{z_0}(z,p)=p$.
Thus, it can now be stated that for any $z$ in $\ZZ[i]$,$\round{r_{\alpha^+}}(z)=\psi_{z_s}(z,\round{r_{\alpha^-}}(z))$ 




\subsection{Sketch of the Algorithm}

With the previous statements, we now have the necessary elements to understand the fundamental ideas of the algorithm. This version of the algorithm simply update a discretized rotation map and call a function that needs to be notified at each update. 
We will later see a more complex version which explains how to make rotate incrementally an image without storing any copy of the original. See Algorithm 1.

Most of the subtilty of the algorithm actually proceeds from the mysteries of the sequence of hinge angles. (See Section \ref{sect:hingenextmyst})
In real implementations, small technical details shall be solved:
Naturally, some additional code has to be added to take in account the fact
that most programs store images into rectangular buffers.
Also, this version of algorithm actually jumps over the hinge angles, and it does not compute the $\rho_m(\alpha)$ for the associated to the hinge angle it self. Of course, this is easily fixable: Apply $\psi$ on the two first quarters, then call the procedure for notification; the algorithm  has then to terminate the transformation associated to the hinge angle by computed the next two quarters, and to call once more the notification procedure. 

\hrule
{\bf Algorithm 1:} incremental\_rotation\_via\_map\_and\_notifications(doextproc)
\hrule
\begin{algorithmic}[1]
\STATE{$\A_\H|_m\leftarrow list\_and\_sort\_hinge\_angles(m)$}
\STATE{$\forall p\in \ZZ[i], R[p] \leftarrow p, (z_s,z_d)\leftarrow beg({\A_\H|_m}(z_s,z_d))$}
\WHILE{true}
\STATE{$(z_s,z_d)\leftarrow next\_spc_{\A_\H|_m}(z_s,z_d)$}
\FOR{$Q \leftarrow 0$ to $3$}
\STATE{k=0}
\WHILE{$|(2k+1)z_s|<m$}
\STATE{$op=R((2k+1)z_si^{Q})$}
\STATE{$np\leftarrow R((2k+1)z_si^{Q})+i^{Q}$}
\STATE{$R((2k+1)z_si^Q)\leftarrow np$}
\STATE{$k\leftarrow k+1$}
\ENDWHILE{}
\ENDFOR{}
\STATE{$doextproc(R,z_s)$}
\ENDWHILE{}
\end{algorithmic}
\hrule
\vspace{0.3em}
{\it $r$ denotes the radius of the image, $op$ means old position, $np$ stands for new position, the function ``doextproc'' is a parameter function that is called each time the image has been set up in configuration that corresponds to the image of a discretized rotation.}
\hrule
\vspace{0.3em}




\subsection{Application to Rotation of Images}
\label{sect:pictrot}

The previous version of the algorithm computes an image of the rotation map,
however the discretized rotation is not intrisically bijective on $\ZZ^2$ and it 
is not suitable to compute incrementally rotations of an image without any copy of the
original.









It is well-known that a point of the discretized rotation has never more
than two antecedents by discrete rotation. Thus, to create lossless discretized rotation, one natural idea  is to store both antecedents when the function is not injective. 
This requires one additional layer: hence, an image in our algorithm shall be an element of $Q^{(\ZZ[i]\times\{0,1\})}$.



Practically, the code is modified such that: the dataspace on which rotation are computed can support up to two ``colors'' for each position of $\ZZ[i]$.
We will use an arbitrary order (the lexical order or anything fast to compute) to decide which pixel will stand on the layer 1.

The details that have been added allows 
to compute the $2$-layers discretized rotation. See Algorithm 2.



\subsection{Analysis of the Complexity}

This last algorithm has really different complexity comparing to the usual algorithm for rotations. This algorithm is a bit slower than the usual one in the general case, but it is faster when incremental rotations are needed. Moreover this algorithm is slowed down by the fact for the moment we cannot assume that there exist a good structure to store the table of hinge angles. While, it is assumed that each processor uses a sine/cosine table which allows to compute sine and cosine very efficiently.

{\it Space Complexity :} 
The second part of the 
algorithm uses the amount of memory
required to store a 2 layer image, plus the rotation map.
Thus, the algorithms uses about 
$3K(m^2)$, (where $K$ is the memory cost to store one color, or one vector (assuming that these two data-types can be stored with the same number of bytes $K$)). The traditional algorithm uses about $2K(m^2)$ bytes of memory.
Thus, for this part the space requirements are of the same order $\O(m^2)$ and similar in terms of multiplicative factors.
The first part of the algorithm requires $\O(m^3log(m))$ bytes of memory 
to construct a $b$-tree of hinge angles in $\A_\H|_m$. This table can be computed once for all.

{\it Time complexity:}
The list-and-sort procedure for hinge angles requires $\O(m^3log(m))$ operations. The time of one iteration of the loop in the second part is intrisically linear in $m$ (the complexity of the user contributed function, doextproc can of course decrease these performances). Although, the main loop is called $\O(m^3)$ times and contains another loop, the algorithm also requires only $\O(m^3)$ operations: we update only $m^2$ pixels and each pixel crosses at most $4m$ times the dual of the grid. To compute all $\rho(\alpha)|_m$, for $\alpha \in [0,2\pi[$ with traditional rotation algorithms would have needed $O(m^5)$ operations.

\vspace{0.3em}

\hrule
{\bf Algorithm 2:} $incremental\_rotation\_of\_an\_image(img_1, doextproc)$
\hrule
\begin{multicols}{2}
\begin{algorithmic}[1]
\STATE{$\A_{\H}|_m\leftarrow list\_and\_sort\_hinge\_angles(m)$}
\STATE{$\forall p\in \ZZ[i], R[p]\leftarrow p, L[p]\leftarrow 1$}
\STATE{$(z_s,z_d)\leftarrow beg({\A_\H|_m}(z_s,z_d))$}
\WHILE{true}
\STATE{$(z_s,z_d)\leftarrow next\_spc_{\A_\H|_m}(z_s,z_d)$}
\FOR{$Q \leftarrow 0$ to $3$}
\STATE{k=0}
\WHILE{$|(2k+1)z_s|<m$}
\STATE{$p\leftarrow (2k+1)z_s$}
\STATE{$op\leftarrow R[p]$}
\STATE{$ol\leftarrow L[p]$}
\STATE{$np\leftarrow op+i^Q$}
\STATE{$img_2(np)\leftarrow img_{ol}(op)$}
\STATE{$R[p]\leftarrow np$; $L[p]\leftarrow 2$; } 

\STATE{$alone\leftarrow true$;}
\FOR{$d \leftarrow 0$ to $3$}
\IF{$R[p+i^d]= op$}
\STATE{$alone \leftarrow false; p'\leftarrow p+i^d$}
\ENDIF
\ENDFOR
\IF{$\Mnon alone$ and $L[p']=2$}
\STATE{$L[p']\leftarrow 1$ }
\STATE{$swap(img_1[op],img_2[op])$}
\ENDIF
\STATE{$alone\leftarrow true$;}
\FOR{$d \leftarrow 0$ to $3$}
\IF{$R[p+i^d]= np$}
\STATE{$alone \leftarrow false; p'\leftarrow p+i^d$}
\ENDIF
\ENDFOR
\IF{$\Mnon$ alone}
\IF{$p'<p$} 
\STATE{$L[p']\leftarrow 1$;$L[p]\leftarrow 2$}
\ELSE{}
\STATE{$L[p']\leftarrow 2$;$L[p]\leftarrow 1$}
\STATE{$swap(img_1[np],img_2[np])$}
\ENDIF{}
\ELSE{}
\STATE{$L[p]\leftarrow 1$}
\STATE{$swap(img_1[np],img_2[np])$}
\ENDIF{}
\ENDWHILE{}
\ENDFOR{}
\STATE{$doextproc(R,z_s,img_1)$}
\ENDWHILE{}
\end{algorithmic}
\end{multicols}
\hrule
\vspace{0.3em}
{\it The incremental discretized rotation algorithm. The notations are the similar to the one used for previous algorithm. Additionally, $img_1$ and $img_2$ denotes the two layers of the picture. $L$ is a part of the configuration that is used to memorize the destination layer of a pixel.}
\hrule

\vspace{0.3em}

%


\subsection{Open-Source Implementation}



We have written an implementation of this algorithm.
The C++ code can be downloaded from \url{http://perso.ens-lyon.fr/bertrand.nouvel/transitive-rotations/}. This implementation relies on similar ideas but is actually slightly different: For historical reasons and other reasons that specific of the future of the code, it uses two types of configurations.
 The first kind is $\rho_\alpha|m$ and is used to compute images by discrete rotation. While $C_\alpha$ (see \cite{medam}) is used for some additional checking.

\section{Complement of specifications}

\subsection{More on Angular Topology of Finite Balls in $\ZZ[i]$}

We denote by $I_{\H}|_m$ the open intervals of angles delimited by two consecutive hinge angles of $\A_{\H}|_m$.  For any angle $\alpha \in \A$ and any $m\in \NN$ either $\alpha \in A_\H|_m$ either there exists a unique $I \in I_\H|_m$ such that $\alpha \in I$. 
We define the function of $Q_m$ from $\A$ to $\A$, such that
for each hinge angle $\alpha \in \A_\H|_m$, $Q_m(\alpha)$ is the angle $\alpha$ itself; and for any angle $\alpha$ that belongs to an interval $I=]\alpha_0, \alpha_1[ \in I_\H|_m$, we define $Q_m(\alpha)$ as the Pythagorean angle which belongs to $I$ which has the smallest radius. We define $\hat{\A}_\H|_m=Q_m(\A)$.
The set $\hat{\A}_\H|_m$ has the following properties:
$\forall \alpha\in \A, \Milexiste \alpha'\in \hat{\A}_\H|_m,\mathrm{such\ that\ } \rho_m(\alpha')=\rho_m(\alpha)$ and $\forall \alpha_1,\alpha_2 \in \hat{\A}_\H|_m, (\alpha_1\neq\alpha_2) \Rightarrow \rho_m(\alpha')\neq \rho_m(\alpha_2)$.

As $\rho_m$ can be seen as a bijection from $\hat{\A}_\H|_m$ to $\rho_m(\A)$, we define the function $\phi_m$ that associates to
each configuration $\rho(\alpha)$ the unique angle $\alpha'\in \hat{\A}_\H|_m$ such that $\rho(\alpha)=\rho(\alpha')$.
Note also that,
if $\alpha, \alpha' \in I=]\alpha_0,\alpha_1[$ and $I \Minter \A_r=\emptyset$ then
$\rho_m(\alpha)=\rho_m(\alpha)$

\def\vtt{\vect{t}}

The hinge angles also have also more properties which are beyond the scope of this paper, and whose proof is mostly arithmetical: If $\alpha$ is a hinge angle then $k\alpha$ is 
a hinge angle for $k\neq\ 0\ mod\ 3$ (by recurrence $k\alpha$ is a hinge angle then $(k+3)\alpha$ is also a hinge angle) , if $\alpha$ is a non flat hinge angle then its sine and its cosine are rationally dependant, $\cround{\ZZ e^{i\alpha}}$ forms span of $2gcd(x,y)$ line of slopes $\frac{x}y$ in $(\RR/\ZZ)[i]$...

Note: The set of hinge angle that has been obtained and studied
here is dependant of the discretization that has been used (and of the real
 center of the rotation).


\subsection{Rotations of the Entire Grid $\ZZ^2$}



If we consider infinite configurations, then $\ZZ[i]^{\ZZ[i]}$ is in bijection with $\A$. More precisely, for any $\alpha,\alpha'\in \A$, $ \alpha={\alpha'}, \round{r_\alpha}=\round{r_\alpha'}$ iff and $\alpha=\alpha'$.
This is trivial since  $lim_{m\mapsto \infty}(\phi_m \circ \rho_m(\alpha))-\alpha\to 0$, .

Moreover, there exists a convergent process that consists computing
gradually each $\rho_m(\alpha)$ with $m$ incrementing through time from 1 to $\infty$.
This process is a well defined procedure to describe discrete rotations.





\subsection{$\epsilon$ - Quasi- Transitivity}




The rotation algorithm specifies which map $\psi$ 
should be applied to transform one rotation map  
$\round{r_{\alpha^-}}(img)$ into another $\round{r_{\alpha^+}}(img)$
 (with respect to the transformations that have previously been applied).
To create a transitive rotation, the $\alpha$-rotation-map of previously applied transformation needs to be embedded within the picture as a part of the data manipulated by the
algorithm\footnote{Of course, it can also be imaginable to record the angle of which the image has been rotated before, but actually we think it is more applications will be interested by a configuration that describes the influence of previously applied transformations on the topology of the space}.

The incremental rotation has been implemented here as an endless process.
The procedure rotates  gradually a picture and
 regularly calls a function to notify the process that the image has 
a new layout. The algorithm can be easily modified to stop when it reaches a specified angle.

From any rotation configuration $C\in ((\ZZ[i])^{\ZZ[i]|_m})$ it is possible to extract the associated slope to the configuration, by computing $\lim_{m\to \infty} tan(\phi_m(\rho_m(\alpha)))$. From the embedded configuration $C=\rho|_m(\alpha_{d_0})$ then we can restart the algorithm from $\alpha_{d_0}=\phi_m(C)$ and then we can rotate incrementally to a new angle $\alpha_{d_1}$. This process can be repeated without any incidence on the precision: the resulting map will stay a discretized rotation map.


If we decide to denote by $\llbracket r_\alpha \rrbracket$, the 
application of the previous algorithm from an angle $\alpha_{d_0}$ (computed
from the embedded configuration), to an angle $\alpha_{d_1}=\alpha_{d_0}+\alpha$.
Then, this operation would be $\epsilon$ - quasi transitive, in the following meaning: $\llbracket r_\alpha \rrbracket \circ \llbracket r_\beta \rrbracket=\llbracket r_{\alpha+\beta+\epsilon(\alpha,\beta)} \rrbracket $ .

More general rules may be found later to apply rotations knowing the input but with broader constraints
 on the admissible type of permutations of $\ZZ[i]$: We can imagine to allow rotations from various centers , translations...



\subsection{Open Question: Efficient Procedure for Computation of the Successor of an Hinge Angle}

\label{sect:hingenextmyst}

One of the remaining problem of our algorithm is that the only way to compute efficiently the successor angle is now, up to our knowledge, 
to construct and use a table of the hinge angles.

However, the question of the structure of hinge angles 
 seems to be linked with some famous 
arithmetic and number theory results. Also, the hinge angles can be seen as the subset of the more general 
set $\A_S$ of angles such that $\alpha$ verifies:
$a cos(\alpha)+b \sin(\alpha)=c,\mathrm{\ with\ } a,b,c \in \{-m,...,m\}$

For our algorithm, we are more interested by the structure induced on 
the generating triples $(p,q,k)$ than by the values of the angles themselves.
 Although the equation seems  familiar due to its similarity to 
formulas for rotations, very few is known about its integer 
solutions.


\section{Conclusions and Perspectives}

This algorithm for discretized rotations has the numerous useful properties:
Since it is an exact algorithm, that is valid on any size of datas,
it returns the same result as the direct discretized rotation of for the associated angle.  
However, this algorithm proceeds incrementally, and it will be useful in all 
procedures that need to do some checking throughout the rotations.
Additionally to be more exact, this algorithm is faster to compute than 
incremental rotations that would iterate the discretized rotation.
More precisely it leads to $\mathcal{O}(n^3)$ updates of pixel instead of
 $\mathcal{O}(n^5)$. 
One of its other property is to be very accurate in terms of spatial error
since the error 
made on the position of an image is 
at most of $\frac{\sqrt{2}}2$ (which is optimal). The fact that it does not use any  sine nor  any cosine strengthen this accuracy.

This work seems to be extendable to the 3-shears rotations.
The main advantage that motivates the use of similar methods to the 3-shears rotations is that 
these rotations are natively bijective.
Therefore, the additional layer becomes obsolete; this would simplify
the algorithm. Some refinements on its associated dynamical system associated 
with hinge angles are certainly required before to can initiate this 
study.

Finally, although the hinge angles are sortable using only 
integer arithmetic, the procedure is slowed down by the use of large integer 
numbers. Thus, to speed up the algorithm, 
as suggested in Section \ref{sect:hingenextmyst}, a trend for ongoing research is to find
an algorithm to return efficiently the successor of an 
hinge angle in $\A_{\H}|_m$.




\nocite*

\bibliography{trxtex.gen.bib}



\end{document}

